\documentclass[preprint,12pt]{elsarticle}
\usepackage{graphicx}
\usepackage{amsfonts,amsmath,amssymb,bm}
\usepackage[dvips,usenames]{color}

\journal{Physica A}

\begin{document}

\begin{frontmatter}



\title{On the non-equivalence of
two standard random walks}


\author[label1]{O. B\'enichou}
\address[label1]{Laboratoire de Physique Th{\'e}orique de la Mati{\`e}re
Condens{\'e}e (UMR CNRS 7600), Universit{\'e} Pierre et Marie Curie (Paris 6) -
4 Place Jussieu, 75252 Paris, France}

\author[label2]{K. Lindenberg}
\address[label2]{Department of Chemistry and
Biochemistry and BioCircuits Institute,
University of California San Diego, La Jolla, CA
92093-0340, USA}

\author[label1]{G. Oshanin}

\begin{abstract}
We focus on two models of nearest-neighbour random walks on $d$-dimensional regular
hyper-cubic lattices that are usually assumed to be identical - the discrete-time
Polya walk, in which the walker steps at each integer moment of time, and the
Montroll-Weiss continuous-time random walk in which the time intervals between successive
steps are independent, exponentially and identically distributed random variables with
mean $1$. We show that while for symmetric random walks both models indeed lead to
identical behaviour in the long time limit, when there is an external bias they lead to
markedly different behaviour.
\end{abstract}


\end{frontmatter}






\section{Introduction}

The random walk is a paradigmatic stochastic process
which underlies a vast variety of phenomena
in physics, chemistry and biology (see, e.g.,~\cite{hugues}).
In particular, two models of lattice random walks are found in the
literature with equal frequency.
The first is the so-called Polya walk (see, e.g.,~\cite{hugues}),
in which the random walker moves  at \textit{discrete} time moments $n = 1,2,\ldots$
on a $d$-dimensional lattice so that
if it is at site ${\bf X'}$ it steps on neighbouring site
${\bf X}$ with probability $p({\bf X})$.
The second model is the Montroll-Weiss random walk~\cite{montroll}, in which the same random walk
evolves in \textit{continuous} time, and the
times $t_j$ between successive jumps are independent,
identically distributed random variables
with a common probability density function $\psi(t)$.
It is often tacitly assumed  that
in the canonical case when the waiting (or pausing) time density is exponential,
i.e., when $\psi(t_j) = \exp\left(-t_j\right)$, the Montroll-Weiss random walk
leads to behaviour \textit{identical} to that of the Polya walk in the limit $t \to \infty$.

The purpose of this short note is to demonstrate
that this is indeed the case for \textit{symmetric} random walks,
but that in the presence of an external bias the two models exhibit markedly
different behaviour.  Although this non-equivalence of the two standard models
can be explained on intuitive grounds,
to the best of our knowledge it has been never
spelled out explicitly.

\section{Polya random walk vs Montroll-Weiss random walk}

Consider a Polya random walk on a $d$-dimensional hyper-cubic lattice with axes
$x$ and $y_j$, $j = 1,\ldots,d-1$, and suppose that there is a constant force $E$ acting
on the walker which points in the positive $x$-direction. In this case, the standard
transition probabilities $p({\bf X})$ which obey detailed balance are given by
\begin{equation}
p_{x} = Z^{-1} e^{\beta E/2}, \,\,\, p_{-x} = Z^{-1} e^{- \beta E/2}, \,\,\, p_{0} = Z^{-1},
\end{equation}
where $\beta$ is the reciprocal temperature, $p_x$, $p_{-x}$, and $p_0$ are the probabilities
to jump in the direction of the applied force, in the direction opposite to the applied force,
and in any direction perpendicular to the force, respectively, and $Z$
is the normalisation constant
\begin{equation}
Z = 2 \cosh\left(\beta E/2\right) + 2 (d - 1).
\end{equation}
Consider the site occupation probability $P_n(x,\{y_j\})$,
that is, the probability of finding the walker at site ${\bf X} = (x,\{y_j\})$ at discrete
time moment $n$.  The characteristic function $\Phi_n(\mathbf{\omega})$ of the site occupation
probability is defined as its discrete Fourier transform,
\begin{equation}
\Phi_n(\mathbf{\omega}) = \sum_{x,\{y_j\} = - \infty}^{\infty} \exp\left(i \omega_x x + i \sum_{j=1}^{d-1} \omega_j y_j\right) P_n(x,\{y_j\}).
\end{equation}
For a Polya walk this function can easily be calculated to give
\begin{equation}
\Phi_n(\mathbf{\omega}) = \left(p_x e^{i \omega_x} + p_{-x} e^{- i \omega_x} + Z^{-1} \sum_{j = 1}^{d - 1} \cos\left(\omega_j\right) \right)^n.
\end{equation}
Differentiating the latter
equation once and twice over $\omega_x$ leads to the following expressions for the first two moments
of the displacement along the $x$-axis:
\begin{equation}
\label{mean1}
\overline{x}_n = \left(p_x - p_{-x}\right) \, n \,,
\end{equation}
and
\begin{equation}
\overline{x^2}_n = \left(p_x - p_{-x}\right)^2 (n - 1) n + \left(p_x + p_{-x}\right) \, n \,,
\end{equation}
so that the variance for motion along the $x$-axis is given by
\begin{equation}
\label{var1}
{\rm Var}_n(x) = \left(p_x \left(1 - p_x\right) + p_{-x} \left(1 - p_{-x}\right) + 2 p_x p_{-x} \right) \, n \,.
\end{equation}
Similarly, differentiating once and twice with respect to any given $\omega_j$
we find that the variance of the site occupation distribution in any direction perpendicular
to the field is given by
\begin{equation}
\label{var11}
{\rm Var}_n(y_j) = Z^{-1} \, n\,.
\end{equation}

To calculate analogous properties for the continuous-time Montroll-Weiss random walk with an
exponential pausing time distribution, we take advantage of the Montroll-Weiss
theorem~\cite{montroll} which
establishes a general relation between $P_n(x,\{y_j\})$ for the discrete-time walk
and its continuous-time
counterpart $\tilde{P}(x,\{y_j\}; t)$ describing the site-occupation probability
for the continuous-time process.

Let $P_{\xi}(x,\{y_j\})$ denote
the generating function of the site occupation probability $P_n(x,\{y_j\})$, i.e.,
\begin{equation}
P_{\xi}(x,\{y_j\}) = \sum_{n=0}^{\infty} P_n(x,\{y_j\}) \, \xi^n \,.
\end{equation}
The Montroll-Weiss theorem states that the Laplace transform of the site-occupation
probability $\tilde{P}(x,\{y_j\};t)$ for the
continuous-time walk and $P_{\xi}(x,\{y_j\})$ are related as follows:
\begin{eqnarray}
\label{relation}
\tilde{P}(x,\{y_j\}; u) &=& \int^{\infty}_0 dt \, \exp(- u t) \, \tilde{P}(x,\{y_j\}; t) \nonumber\\
&=& \frac{1 - \phi(u)}{u} \, P_{\psi(u)}(x,\{y_j\}) .
\end{eqnarray}
Here $\phi(u)$ is the Laplace transformed pausing-time density.
Multiplying both sides of Eq.~(\ref{relation}) by $x$ and $x^2$, and summing over all lattice sites,
we readily find the following results for the moments of the displacement
of the continuous-time random walk along the $x$-axis:
\begin{equation}
\label{mean2}
\overline{x}_t = \left(p_x - p_{-x}\right) \, t \,,
\end{equation}
and
\begin{equation}
\label{var2}
{\rm Var}_t(x) = \left(p_x + p_{-x}\right) \, t \,.
\end{equation}
In a similar fashion we find that
the variance in any direction perpendicular to the field obeys
\begin{equation}
\label{var22}
{\rm Var}_t(y_j) = Z^{-1} \, t \,.
\end{equation}

\begin{figure}[ht]
  \centerline{\includegraphics*[width=0.55\textwidth]{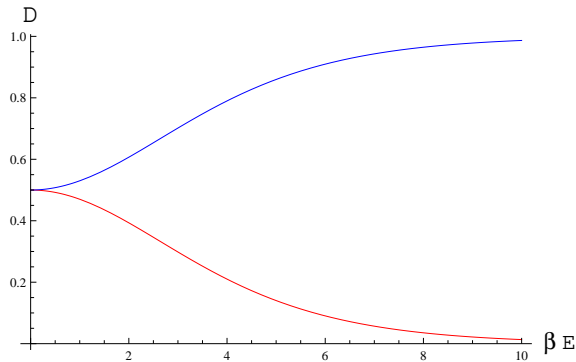}}
  \caption{(Color online)
The reduced variance $D$ along the $x$-axis as a function of $\beta E$ 
for two-dimensional lattices. The red
line is the result for the Polya walk, $D = {\rm Var}_n(x)/n$, Eq.~(\ref{var1}), while the blue
line is the result for the continuous-time random walk, $D = {\rm Var}_t(x)/t$, Eq.~(\ref{var2}).}
\label{fig0}
\end{figure}

\section{Conclusions}

We conclude with the comparison of the results in Eqs.~(\ref{mean1}) and (\ref{mean2}),
Eqs.~(\ref{var1}) and (\ref{var2}), and Eqs.~(\ref{var11}) and (\ref{var22}). One
immediately notices that  both models predict exactly the same values for the
``velocities" $\bar{x}_n/n$ and $\bar{x}_t/t$ (first moments of the site occupation
distributions), as well as for the reduced vriances Var$_n(y_j)/n$ and Var$_t(y_j)/t$
obtained from these distributions in the direction perpendicular to
the field. On the contrary, for
these two models, often assumed to lead to identical results, the
diffusion coefficients (variances) along
the $x$-axis exhibit markedly different behaviours as depicted in Fig.\ref{fig0}.
The variance along the $x$-direction for the
Polya walk is a monotonically \textit{decreasing} function of the field,
while for the continuous-time random walk with exponential waiting time distribution it
is (for $d \geq 2$) a monotonically \textit{increasing} function of the field which saturates
at a finite value as $\beta E \to \infty$. In one-dimension the variance
is \textit{independent} of the value of the applied field. This difference in behaviours
stems from the fact that for the discrete-time random walk the number of steps does not
fluctuate so that the system tends toward a deterministic ballistic motion as
$\beta E \to \infty$, that is, all realisations in this limit cover the same ground and the variance
vanishes.  On the other hand, for the continuous-time
random walk the number of steps fluctuates from realisation to realisation, which
contributes to the variance no matter how strong the field
(however, only along the $x$-direction). Note that such a striking difference between these two models is completely
lost, of course, when one turns to the continuous-space and time description in the diffusion limit.
Note as well that for $\beta E \ll 1$, both results show the same dependence
on the field, which ensures the validity of the Einstein relation between
diffusion coefficient and mobility for both models.


\vspace{2pc}


\begin{thebibliography}{99}

\bibitem{hugues} Hughes B D (1995) {\it  Random Walks and Random Environments}, (Oxford University Press, Oxford).
 %
\bibitem{montroll} Montroll E W and Weiss G H (1965) J. Math. Phys. {\bf 6} 167.
%
\end{thebibliography}
\end{document}